\documentclass[twocolumn,twocolappendix,appendixfloats]{aastex6}

\usepackage{subfigure}
\usepackage{threeparttable}
\usepackage{multirow}
\usepackage{remreset}
\usepackage{CJKutf8}

\collaborationName{Friends of AASTeX}
\shorttitle{saturation effect on BAL variability}
\shortauthors{Lu \& Lin}
\begin{document}


\title{Saturation effect on photoionization-driven broad absorption line variability}


\author{
\begin{CJK*}{UTF8}{gbsn}
Wei-Jian Lu (陆伟坚)\altaffilmark{1} and Ying-Ru Lin (林樱如)\altaffilmark{2}
\end{CJK*}
}
\affil{School of Information Engineering, Baise University, Baise 533000, China}

\altaffiltext{1}{E-mail: william\_lo@qq.com (W-J L)}
\altaffiltext{2}{E-mail: yingru\_lin@qq.com (Y-R L)}



\begin{abstract}
We study the saturation effect on broad absorption line (BAL) variability through {a variation phenomenon, which shows significant variation in Si\,{\footnotesize IV} BAL but no, or only small, change in C\,{\footnotesize IV} BAL (hereafter Phenomenon I). First, we explore a typical case showing Phenomenon I, quasar SDSS J153715.74+582933.9 (hereafter J1537+5829).} We identify four narrow absorption line (NAL) systems within its Si\,{\footnotesize IV} BAL and two {additional} NAL systems within its C\,{\footnotesize IV} BAL, and confirm their coordinated weakening. Combining with the obvious strengthening of the ionizing continuum, we attribute the BAL variability in J1537+5829 to the ionization changes caused by the continuum variations. Secondly, a statistical study based on multiobserved quasars from SDSS-I/II/III is presented. We confirm that (1) the moderate anticorrelation between the fractional variations of Si\,{\footnotesize IV} BALs and the continuum in 74 quasars that show Phenomenon I and {(2) the sample showing BAL variations tends to have larger ionizing continuum variations. These results reveal the ubiquitous effect of the continuum variability on Phenomenon I and BAL variation.} We attribute the {relative} lack of variation of C\,{\footnotesize IV} BALs in Phenomenon I to the saturation effects. Nonetheless, these absorbers are not very optically thick {in Si\,{\footnotesize IV}} and the ionization changes in response to continuum variations could be the main driver of their variations. Finally, we find that the saturation effect on BAL variability can well explain many phenomena of BAL variations that have been reported before.
\end{abstract}




\section{Introduction} \label{sec:intro}
\defcitealias{Lu2018a}{LLQ2018}
{At present, it is clear that both transverse motion (TM) and ionization change (IC) of the  outflow cloud can be mechanisms that drive the variability of broad absorption line (BAL). Both of these two sources of variability have observational evidence. 
For the TM scenario, early works have not found an obvious relation between BALs and the continuum variability \citep{Gibson2008,Vivek2014,Wildy2014}; in recent works, the strength of P\,{\footnotesize V} in BAL outflows inferred that the BALs are very saturated with large total column densities, which may offer an evidence for the TM model \citep{Capellupo2017,Moravec2017, McGraw2018}. For the IC scenario,} coordinated multitrough variability (e.g., \citealp{Capellupo2012,Capellupo2013,Filiz2013,Wildy2014,Wang2015}) and the coordinated variability between BALs and the ionizing continuum \citep{Wang2015} have been found, {these results tend to support the IC scenario}. {For instance, \citet{Filiz2013} estimated that (56$\pm$7)\% of trough variations are arising from a mechanism correlated between troughs, such as ICs. Besides, \citet{Wang2015} showed that BAL gas can, in principle, show large changes in response to only small continuum variations.} More recently, based on the multiepoch observed BAL quasars from SDSS-I/II/III, \citet{He2017} have estimated statistically that at least 80\% of the BAL variations are mainly driven by variations of the ionizing continuum. Based on the same data sample of \citet{He2017}, Lu et al. (\citeyear{Lu2018a}, hereafter \citetalias{Lu2018a}) have revealed the moderate anticorrelations with a high significance level between fractional equivalent width (EW) variations ($\Delta \rm EW/\langle EW \rangle$) of BALs (for both Si\,{\footnotesize IV} and C\,{\footnotesize IV} BALs) and the fractional flux variations of the continuum ($\Delta \rm F_{\rm cont}/\langle \rm F_{cont} \rangle$). 

\begin{figure*}
\includegraphics[width=2.1\columnwidth]{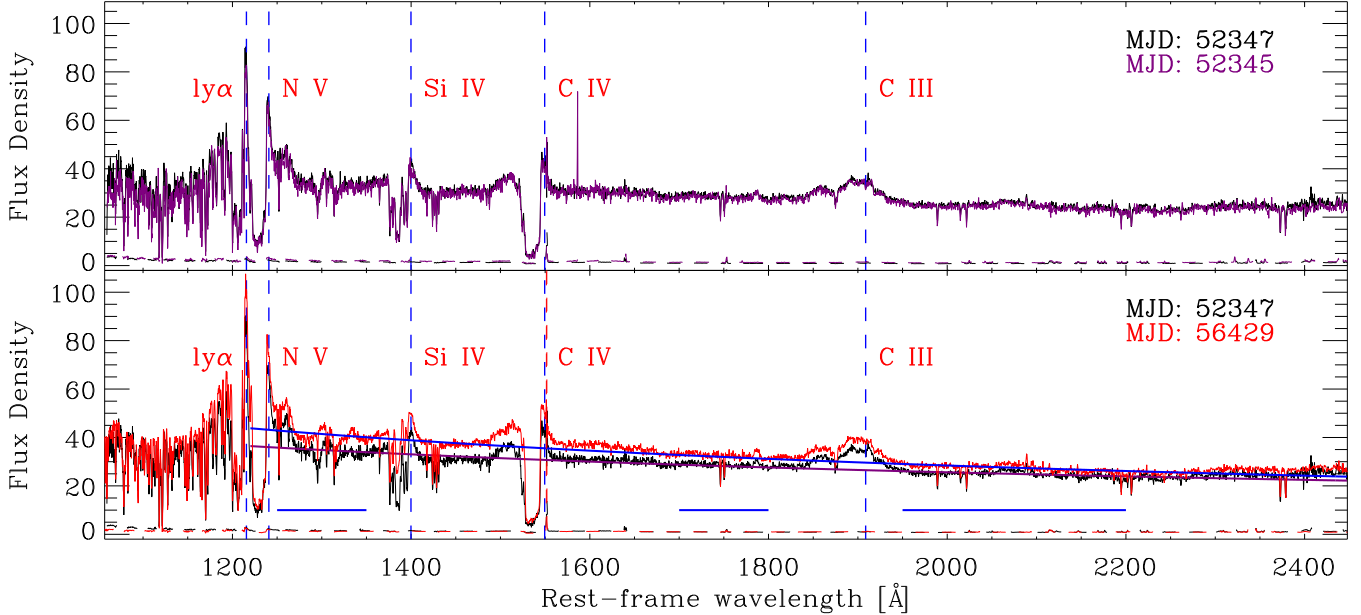}
\caption{Three spectra of J1537+5829. The SDSS MJDs of the spectra are labeled. The flux density is in units of $\rm 10^{-17}~erg~s^{-1}~cm^{-2}$. {The dashed lines shown near the bottom of each panel are the formal 1$\sigma$ errors.} The blue vertical dashed lines mark out the main emission lines. The blue and purple lines in the bottom panel are the power-law continua for spectra of J1537+5829 on MJD 52347 and 56429. {The blue horizontal bars below the spectra are the regions used to fit the power-law continua. }  \label{fig.1}}
\end{figure*}

Previous studies have found that BALs are frequently saturated but not black, which suggests that these absorbers do not completely obscure the continuum emitter (e.g., \citealp{Hamann1998,Arav1999a}).
Saturation of BALs has crucial influence on the exploration of BAL variability, {because the EW measurements of saturated BALs cannot accurately reflect the true value of optical depths and column densities (e.g., \citealp{Hamann1998,Arav1999a,Gabel2003}).} BALs do vary from days to years (e.g., \citealp{Filiz2013,Grier2015}), but they can only weakly respond to the fluctuations of the ionizing continuum when the BAL troughs are highly saturated. Namely, a severely saturated C\,{\footnotesize IV} BAL will disfavor the IC scenario as the cause of its variability. If a severely saturated BAL still shows time variation, then it tends to support the TM scenario as its variability cause. Thus, aiming to confirm the cause of BAL variability, some authors have attempted to prove the saturation of BAL systems (e.g., \citealp{Arav1999b}) at first through, for example, the detection of P\,{\footnotesize V} $\lambda\lambda$1118, 1128 BALs (e.g., \citealp{Hamann1998,Capellupo2014,Capellupo2017,Moravec2017}). 

{Since the BALs are frequently suffering from saturation, how does the saturation feature affect the BAL variation? What is the physical mechanism behind the variation of BALs that show the saturation feature? 
In this paper, we study a variation phenomenon that shows significant variation in Si\,{\footnotesize IV} BAL but no, or only small, change in C\,{\footnotesize IV} BAL (hereafter Phenomenon I).} This interesting signature of BAL variation may offer clues to the above questions. {We first give an analysis of a typical case showing Phenomenon I, quasar SDSS J153715.74+582933.9 ($z_{\rm em}=2.595$; \citealp{Paris2017}; hereafter J1537+5829). Then we present a statistical study of the main driver of the variations of the BALs that show Phenomenon I, based on quasars multiobserved by SDSS-I/II/III.} 
The paper is structured as follows. {Section \ref{sec:J1537} presents the study of J1537+5829. Section \ref{sec:statistical} contains the details of the statistical analysis. Section \ref{sec:implication} discusses the implications of saturation effect, and our conclusions are provided in section \ref{sec:conclu}.}

\begin{figure*}
\includegraphics[width=2.1\columnwidth]{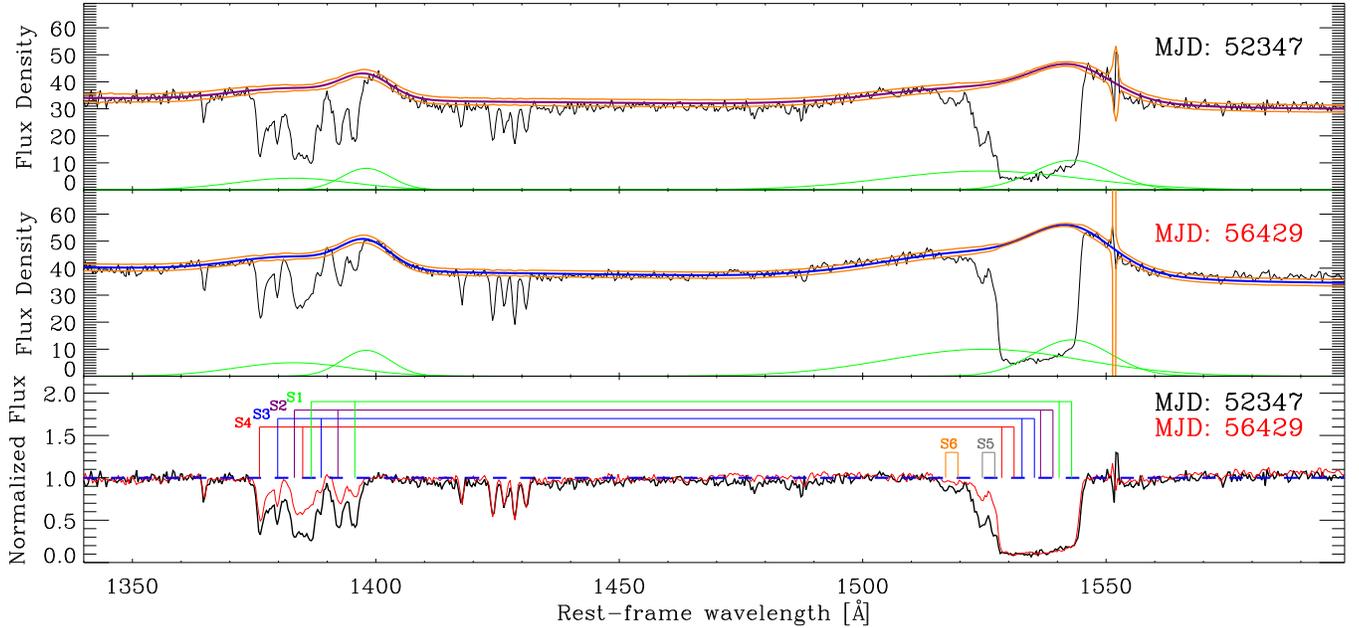}
\caption{Pseudo-continuum fits (top and middle panels) and pseudo-continuum normalized spectra (bottom panel) of J1537+5829 on MJD 52347 and 56429. The flux density is in units of $\rm 10^{-17}~erg~s^{-1}~cm^{-2}$. The purple line in the top panel and the blue line in middle panel are the pseudo-continua, the orange lines are the pseudo-continua that have added the flux uncertainties. {The green Gaussian profiles in the bottom of the top and middle panels are the Gaussian components used to fit the Si\,{\footnotesize IV} and C\,{\footnotesize IV} emission lines.} The green, purple, blue, red, gray and orange lines in the bottom panel mark out the six identified NAL systems within the Si\,{\footnotesize IV} and C\,{\footnotesize IV} BALs. 
\label{fig.2}}
\end{figure*}
\begin{figure*}
\includegraphics[width=2.1\columnwidth]{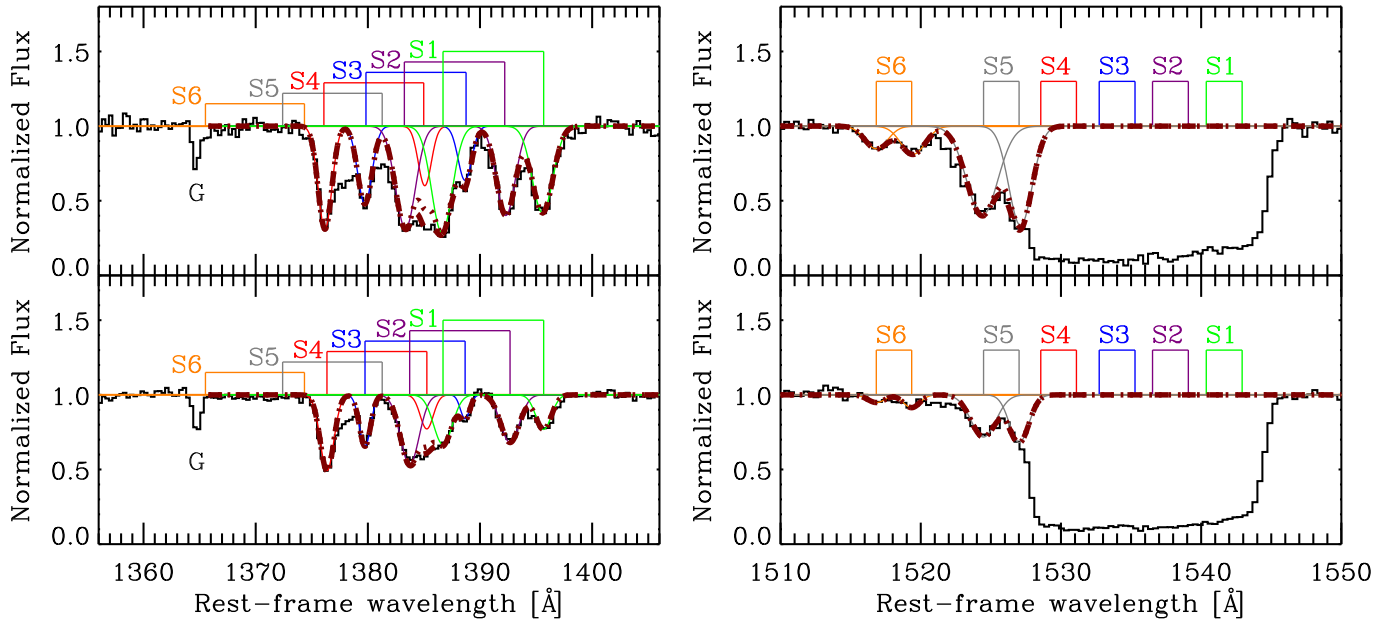}
\caption{NAL absorption systems within the BALs in J1537+5829 (left panel: four NAL absorption systems (S1$\sim$S4) within the Si\,{\footnotesize IV} BAL trough; right panel: two NAL absorption systems (S5 and S6) within the C\,{\footnotesize IV} BAL trough). The top and bottom spectra are snippets from the MJD 52347 and 56429 normalized spectra, respectively. {The brown broken lines and brown dotted lines represent the additive fit and multiplicative fit models, respectively.  ``G" marks a strong Galactics absorption line.} \label{fig.3}}
\end{figure*}


\section{{A typical case showing Phenomenon I}} \label{sec:J1537} 

\defcitealias{Lu2018b}{LL2018}

Currently, Lu \& Lin (\citeyear{Lu2018b}, hereafter \citetalias{Lu2018b}) have discovered that some of the BALs (Type N BALs\footnote{{This type of BALs were referred to ``Type II BALs" in \citetalias{Lu2018b}. To avoid potentially confusing ``Type II BAL quasars" with narrow-line AGNs, which exhibit BAL troughs, we change the term to ``Type N BALs" in this paper. 
Similarly, we rename the ``Type I BALs," which show relatively smooth BAL troughs that cannot be decomposed into multiple NALs, as ``Type S BALs."}}) are actually a complex of NALs, and that the decomposition of a Type N BAL into NALs can serve as an effective way to study the BAL outflows. In this section, we present the analysis of a Type N BAL quasar J1537+5829, which shows Phenomenon I.
\subsection{Spectroscopic analysis} \label{sec:spectro} 

J1537+5829 was observed by the Sloan Digital Sky Survey (SDSS; \citealp{York2000}) on MJD 52345 and 52347, and by the Baryon Oscillation Spectroscopic Survey (hereafter ``BOSS"; \citealp{Dawson2013}) on MJD 56429. The SDSS spectra cover a wavelength range between $\sim$3800 and 9200\,\AA~and have a resolution ranging from $\sim$1850 to 2200, while the BOSS spectrum covers a wavelength from $\sim$3600 to 10 000\,\AA~and has a resolution within $\sim$1300--3000. The median signal-to-noise ratio (S/N) of the MJD 52345, 52347, and 56429 spectra of J1537+5829 are 16.75, 23.73, and 30.81 per pixel, respectively. In this section, we choose the last two spectra to study because  (1) the time interval between the first two observations is just 2 days, and thus these two spectra are almost the same (Figure \ref{fig.1}) and (2) the last two spectra have higher S/Ns than the first one.

As a BAL quasar, J1537+5829 has been studied in some systematic BAL studies \citep{Trump2006,Gibson2009a,Scaringi2009,Allen2011,Bruni2014,Filiz2014,He2015, He2017}. The MJD 52347 spectrum of J1537+5829 has a balnicity index (BI; see the definition in \citealp{Weymann1991}) of 455.2 and 429.2\,$\rm km\,s^{-1}$ for the Si\,{\footnotesize IV} and C\,{\footnotesize IV} BALs (these estimations are taken from \citealp{Allen2011}), respectively. As Type N BALs, both the Si\,{\footnotesize IV} and C\,{\footnotesize IV} BALs in J1537+5829 actually contain multiple NAL systems, but they show different trough profiles: Si\,{\footnotesize IV} BAL shows multiple-absorption troughs, while the C\,{\footnotesize IV} BAL has a P-Cygni/trough-like shape.

From the improved flux calibration of the SDSS data release 14 (DR14; \citealp{Abolfathi2018}), we downloaded the spectra of J1537+5829. We then fitted iteratively the power-law continuum for each spectrum based on several relatively line-free (RLF) wavelength regions (1250--1350, 1700--1800, 1950--2200, 2650--2710\,\AA~ in the rest frame; defined by \citealp{Gibson2009a}). In order to reduce the influences of emission/absorption lines and/or remaining sky pixels, we removed the pixels that are outside 3$\sigma$ significance during the fitting. The power-law continuum fits of the spectra are shown in Figure \ref{fig.1}.

Figure \ref{fig.1} shows that both the Si\,{\footnotesize IV} and C\,{\footnotesize IV} BALs in J1537+5829 are superimposed on the corresponding emission lines. To measure the absorption lines more accurately, we fit the Si\,{\footnotesize IV} and C\,{\footnotesize IV} emission lines using Gaussian profiles. 
{For the Si\,{\footnotesize IV} emission line, it fits well with two Gaussian components with full widths at half maximum (FWHMs) of 2525 and 6125\,$\rm km\,s^{-1}$ 
and velocity offsets of 386 and 3620\,$\rm km\,s^{-1}$, respectively. 
For the C\,{\footnotesize IV} emission line, it fits well with two Gaussian components with FWHMs of 3660 and 9258\,$\rm km\,s^{-1}$ 
and velocity offsets of 1256 and 4774\,$\rm km~s^{-1}$, respectively.}
Combining the power-law continuum and the Gaussian profiles for emission lines, we got the final pseudo continuum for each spectrum (Figure \ref{fig.2}). Then we measured the absorption lines in the pseudo-continuum normalized spectra.

{Following \citetalias{Lu2018b}, we did the additive fit to the complex NALs within the Si\,{\footnotesize IV} BAL trough using four pairs of Gaussian functions (left panel of Figure \ref{fig.3}). {We also overplotted the spectrum resulting from multiplying together the residual fluxes from the additive fit as brown dotted lines.} We demonstrate that the difference between the additive spectrum and the multiplicative spectrum is small in most of the wavelengths. But we note that the red component of S4 on MJD 52347 spectrum may suffer from a little underestimation according to the multiplicative spectrum.} 
We also note that we can only identify the lower limit of the numbers of the absorption systems due to the low resolution of the SDSS/BOSS spectra and the blending of the  absorption lines. For instance, there is absorption between the blue components of S3 and S4 in both epoches, which means that four Gaussian NALs are not sufficient to match the full Si\,{\footnotesize IV} BAL in this object. One may infer partial covering of the outflow from the ``nonblack saturation" of C\,{\footnotesize IV} BALs. However, the low resolution of the SDSS/BOSS spectra also does not allow us to correctly measure the coverage fraction of NALs, we therefore accepted a full coverage during our line fitting. Although suffering from severe blending, two C\,{\footnotesize IV} NAL doublets at relatively higher outflow velocities can still be identified (the right panel of Figure \ref{fig.3}). 

We measured the velocity, EW, FWHM, and fractional EW variation of the absorption lines based on the Gaussian fits. The methods for calculating the EWs of the absorption line are the same as those of \citetalias{Lu2018b} (see Equations (1), (2) and (3) of \citetalias{Lu2018b}), and the methods for calculating the fractional EW variations are the same as those of \citetalias{Lu2018a} (Equation (2) of \citetalias{Lu2018a}). Measurements of the absorption lines are listed in Table \ref{tab.1}. 

\begin{table*}[h]
    \centering
\caption{Measurements of Si\,{\footnotesize IV} and C\,{\footnotesize IV} BALs \label{tab.1}}
\begin{tabular}{lcccccccc} 
\hline 
\hline 
Species & $z_{\rm abs}$ & Velocity & \multicolumn2c{MJD:52347} & \multicolumn2c{MJD:56429} & Fractional  &Note\\
\cline{4-7}
 & & &EW &FWHM &EW &FWHM &EW &    \\
 & &($\rm km~s^{-1}$) & (\AA) & ($\rm km~s^{-1}$) & (\AA) & ($\rm km~s^{-1}$) & variation & \\
\hline
Si\,{\footnotesize IV}$\lambda$1393	&	\multirow{2}*{	2.5497	}	&	\multirow{2}*{	3805 			}	&	$	1.07 	\pm	0.03	$	&	314 	&	$	0.79	\pm	0.03	$	&	314 	&	$	-0.15 	\pm	0.046 	$	&	\multirow{2}*{	S4	}	\\
Si\,{\footnotesize IV}$\lambda$1402	&				&						&	$	0.59 	\pm	0.04	$	&	298 	&	$	0.34	\pm	0.06	$	&	298 	&	$	-0.27 	\pm	0.062 	$	&		 		\\
Si\,{\footnotesize IV}$\lambda$1393	&	\multirow{2}*{	2.5588	}	&	\multirow{2}*{	3031 			}	&	$	0.84 	\pm	0.04	$	&	327 	&	$	0.44	\pm	0.04	$	&	256 	&	$	-0.31 	\pm	0.052 	$	&	\multirow{2}*{	S3	}	\\
Si\,{\footnotesize IV}$\lambda$1402	&				&						&	$	0.56 	\pm	0.05	$	&	311 	&	$	0.17	\pm	0.07	$	&	212 	&	$	-0.53 	\pm	0.064 	$	&		 		\\
Si\,{\footnotesize IV}$\lambda$1393	&	\multirow{2}*{	2.5682	}	&	\multirow{2}*{	2246 			}	&	$	1.54 	\pm	0.03	$	&	454 	&	$	0.98	\pm	0.04	$	&	426 	&	$	-0.22 	\pm	0.038 	$	&	\multirow{2}*{	S2	}	\\
Si\,{\footnotesize IV}$\lambda$1402	&				&						&	$	1.26 	\pm	0.04	$	&	423 	&	$	0.67	\pm	0.06	$	&	423 	&	$	-0.31 	\pm	0.042 	$	&		 		\\
Si\,{\footnotesize IV}$\lambda$1393	&	\multirow{2}*{	2.5766	}	&	\multirow{2}*{	1542 			}	&	$	1.65 	\pm	0.03	$	&	467 	&	$	0.7	\pm	0.05	$	&	425 	&	$	-0.40 	\pm	0.037 	$	&	\multirow{2}*{	S1	}	\\
Si\,{\footnotesize IV}$\lambda$1402	&				&						&	$	1.34 	\pm	0.04	$	&	464 	&	$	0.41	\pm	0.07	$	&	352 	&	$	-0.53 	\pm	0.041 	$	&		 		\\
C\,{\footnotesize IV}$\lambda$1548	&	\multirow{2}*{	2.5222	}	&	\multirow{2}*{	6136 			}	&	$	0.31 	\pm	0.18	$	&	378 	&	$	0.08	\pm	0.64	$	&	285 	&	$	-0.59 	\pm	0.084 	$	&	\multirow{2}*{	S6	}	\\
C\,{\footnotesize IV}$\lambda$1551	&				&						&	$	0.42 	\pm	0.17	$	&	401 	&	$	0.13	\pm	0.56	$	&	271 	&	$	-0.53 	\pm	0.073 	$	&		 		\\
C\,{\footnotesize IV}$\lambda$1548	&	\multirow{2}*{	2.5397	}	&	\multirow{2}*{	4647 			}	&	$	1.70 	\pm	0.04	$	&	515 	&	$	0.34	\pm	0.07	$	&	412 	&	$	-0.67 	\pm	0.036 	$	&	\multirow{2}*{	S5	}	\\
C\,{\footnotesize IV}$\lambda$1551	&				&						&	$	1.41 	\pm	0.03	$	&	386 	&	$	0.21	\pm	0.05	$	&	322 	&	$	-0.74 	\pm	0.040 	$	&		 		\\
C\,{\footnotesize IV} BAL	&		...		&		$-4283	\sim	-674^{\rm a}	$	&	$	14.94 	\pm	1.22	$	&	$3609^{\rm b}$ 	&	$	14.62	\pm	0.62	$	&	$3609^{\rm b}$ 	&	$	-0.01 	\pm	0.007 	$	&		S1$\sim$S4		\\
C\,{\footnotesize IV} BAL	&		...		&		$-6785	\sim	-674^{\rm a}	$	&	$	18.55 	\pm	1.77	$	&	$6111^{\rm b}$ 	&	$	16.09	\pm	0.95	$	&	$6111^{\rm b}$ 	&	$	-0.07 	\pm	0.006 	$	&		S1$\sim$S6		\\
Si\,{\footnotesize IV} BAL	&		...		&		$-5576	\sim	-279^{\rm a}	$	&	$	10.15 	\pm	1.73	$	&	$5297^{\rm b}$ 	&	$4.84\pm0.95$	&	$5297^{\rm b}$ 	&	$	-0.35 	\pm	0.010 	$	&		S1$\sim$S4		\\

\hline 
\end{tabular}
\begin{tablenotes}
\footnotesize
\item$^{\rm a}$Velocity range of the BAL troughs with respect to the emission rest frame.
\item$^{\rm b}$Total width calculated from edge-to-edge of the BAL troughs.
\end{tablenotes}
\end{table*}

\subsection{Evidence for IC} \label{subsec:ic} 
We hold the view that the mechanism responsible for the absorption line variability in J1537+5829 is the IC scenario for two reasons: (1) coordinated multiple-trough weakening and (2) obvious continuum strengthening. For the first reason, well coordinated variations between multiple NAL/BAL systems (e.g., \citealp{Hamann2011,Chen2015,Wang2015,McGraw2017}) or over a large velocity interval of a BAL trough (e.g., \citealp{Grier2015,McGraw2017,McGraw2018}) can serve as strong evidence supporting the IC scenario to explain absorption variation. The phenomenon of coordinated absorption line variability is difficult to explain through the TM scenario because it would require coordinated motions of many distinct outflow structures \citep{Misawa2005,Hamann2011}. In the case of J1537+5829, the coordinated weakening among different NAL absorption systems (S1$\sim$S4) within the Si\,{\footnotesize IV} BAL trough is detected (see Figure \ref{fig.2}, Figure \ref{fig.3} and Table \ref{tab.1}). Although belonging to the same systems with the Si\,{\footnotesize IV} NAL components, the S1$\sim$S4 NAL components within the C\,{\footnotesize IV} BAL trough show no obvious variation between the two observations. We ascribe this phenomenon to the saturation in C\,{\footnotesize IV}. As shown in Figure \ref{fig.2} and Table \ref{tab.1}, the NAL components (S1$\sim$S4) of C\,{\footnotesize IV} show larger EWs than Si\,{\footnotesize IV} and are blended severely, resulting in a typical trough-like profile but not black. This suggests that these C\,{\footnotesize IV} NAL components (S1$\sim$S4) are saturated and {do not completely obscure the emission regions at these wavelengths (continuum emission region and C\,{\footnotesize IV} broad emission line region).} However, the other two absorption systems (S5 and S6) within the C\,{\footnotesize IV} BAL, which are relatively weaker at higher outflow velocities, do show coordinated weakening. Thus we speculate that the coordinated ionization state variations have occurred in the saturated NAL components (S1$\sim$S4) within the C\,{\footnotesize IV} BAL. 

For the second reason, anticorrelations between the variations of the ionizing continuum and UV outflow lines have recently been proved (\citealp{Lu2017}; \citetalias{Lu2018a}), which can serve as evidence for the idea that the IC scenario is the primary driver of UV absorption trough variability. Thus, the continuum variation can reveal the cause of UV absorption line variability, to some extent. In the case of J1537+5829, while the absorption lines show coordinated weakening between the two observations, the power-law continuum shows a fractional enhancing of 0.12$\pm$0.025 (see also Figure \ref{fig.1}). This is {compatible} with the previous anticorrelation. Based on the above analysis, we ascribe the absorption line variability in J1537+5829 to the IC scenario, which is the response to the variation of the ionizing continuum.

Based on the case of J1537+5829, we point out that the saturation signature of a  C\,{\footnotesize IV} BAL does not always mean that the TM scenario is the driver of the variation of this BAL system. Although most areas in the saturated C\,{\footnotesize IV} BAL in J1537+5829 lack variability, the coordinated ionization variations of the outflow structures are confirmed due to the coordinated variation of Si\,{\footnotesize IV} NALs at the corresponding velocities.

\subsection{Comparison with \citetalias{Lu2018b}} \label{subsec:compar} 
As another case of showing Type N BALs, J1537+5829 has many similar qualities compared to the quasar example, J002710.06--094435.3 (hereafter J0027--0944), in \citetalias{Lu2018b}. First and foremost, both the Si\,{\footnotesize IV} and C\,{\footnotesize IV} BALs consist of multiple NAL components that show coordinated time variations for both of these two quasars. In fact, the discussions of J0027--0944 in \citetalias{Lu2018b}, on the origin and variability cause of the BALs (section 3.1 of \citetalias{Lu2018b}), the constraints on the inclination model (section 3.2 of \citetalias{Lu2018b}), the explanation for different profile shapes of BALs (section 3.3 of \citetalias{Lu2018b}) and the clumpy structure of outflows (section 3.4 of \citetalias{Lu2018b}), all apply equally to J1537+5829. The biggest difference between these two sources is that the C\,{\footnotesize IV} BAL shows global variation in J0027--0944,  while only small partial change in J1537+5829. However, we do not think that these two C\,{\footnotesize IV} BALs are fundamentally different. The essential reason for this difference is that the C\,{\footnotesize IV} BAL of J1537+5829 suffers from more saturation than that of J0027--0944. Through the cases of J1537+5829 and J0027--0944, we find that the decomposition of a BAL into NALs can serve as an effective way to determine the mechanism of absorption line variability. In fact, the essence of the global variation of a Type N BAL is the coordinated variability of multiple NALs. 

{The phenomenon of the NAL complex as a form of BALs reveals that the line of sight intersects several physically separated outflow components. If the Type N BALs is universal, then it may serve as evidence that the outflow is clumpy in physical space, which, to some extent, supports the schematic of inhomogeneous partial coverage} (IPC; \citealp{Hamann2001,deKool2002,Hamann2004,Arav2005,Sabra2005}). More significantly, recent numerical simulations by \citet{Matthews2016} show that the ionization state can be moderated sufficiently at more realistic X-ray luminosities when incorporating clumping with a filling factor of $\sim$0.01. Namely, clumpy structure of outflow may be self-shielded and thus may offer another explanation to solve the overionization problem (see also \citealp{Hamann2013}). Later systematic study of Type N BALs will {quantitatively} determine the universality of Type N BALs in outflows (Lu et al., in preparation), which may provide more insights into the above discussion.

\section{statistical analysis} \label{sec:statistical} 
{To investigate the main driver of the BAL variations in cases like J1537+5829 showing Phenomenon I, we present a statistical analysis based on the BAL sample selected from SDSS-I/II/III.}

\subsection{Sample selection}\label{subsec:tm}
{Our multiobserved BAL sample was derived from the catalog provided by \citet{He2017}. This catalog contains 9918 spectrum pairs in 2005 BAL quasars (hereafter the Total sample) with a redshift range of $1.9 \textless z \textless 4.7$ and an S/N level of S/N $\textgreater$10 in at least one observation. First of all, we downloaded these spectra from SDSS DR14 and fitted the power-law continuum for each of them using the same procedure described in \citetalias{Lu2018a}. The strength of the ionizing continuum of each spectrum was estimated using the power-law continuum flux at 1450\,\AA. Then we only keep quasars that meet two criteria: (1) the Si\,{\footnotesize IV} BAL shows significant variation (with a confidence level of $\Delta \rm W \textgreater5\sigma$\footnote{See Equation (1) of \citetalias{Lu2018a} for the definition of $\Delta W$.}) and (2) the absolute value of fractional EW variation of C\,{\footnotesize IV} BAL less than 0.08 (this threshold is chosen  referring to the fractional C\,{\footnotesize IV} BAL variation in J1537+5829). Finally, including  J1537+5829, we found 318 spectrum pairs in 74 quasars (hereafter the Phenomenon I sample) meet these criteria.} 

\begin{figure}
\includegraphics[width=1.01\columnwidth]{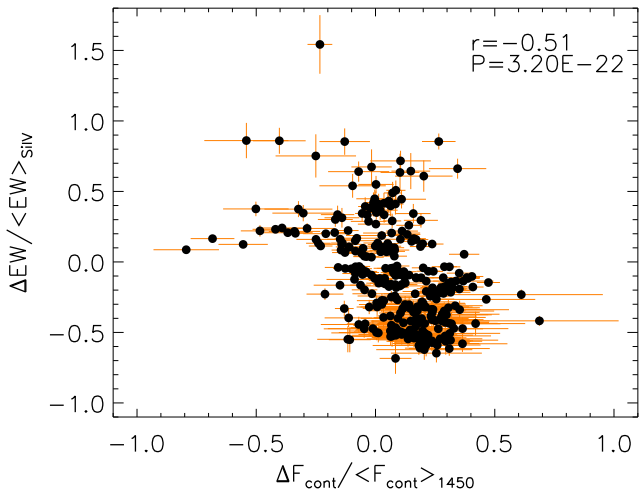}
\caption{Plots of fractional EW variations of Si\,{\footnotesize IV} BALs and fractional flux variations of the continuum. The median error values of $\rm \Delta EW/\langle EW \rangle_{Si\,{\footnotesize IV}}$ and $\Delta \rm F_{\rm cont}/\langle \rm F_{cont} \rangle_{1450}$ are 0.040 and 0.082, respectively.
\label{fig.4}}
\end{figure}

\begin{figure*}
\includegraphics[width=2.1\columnwidth]{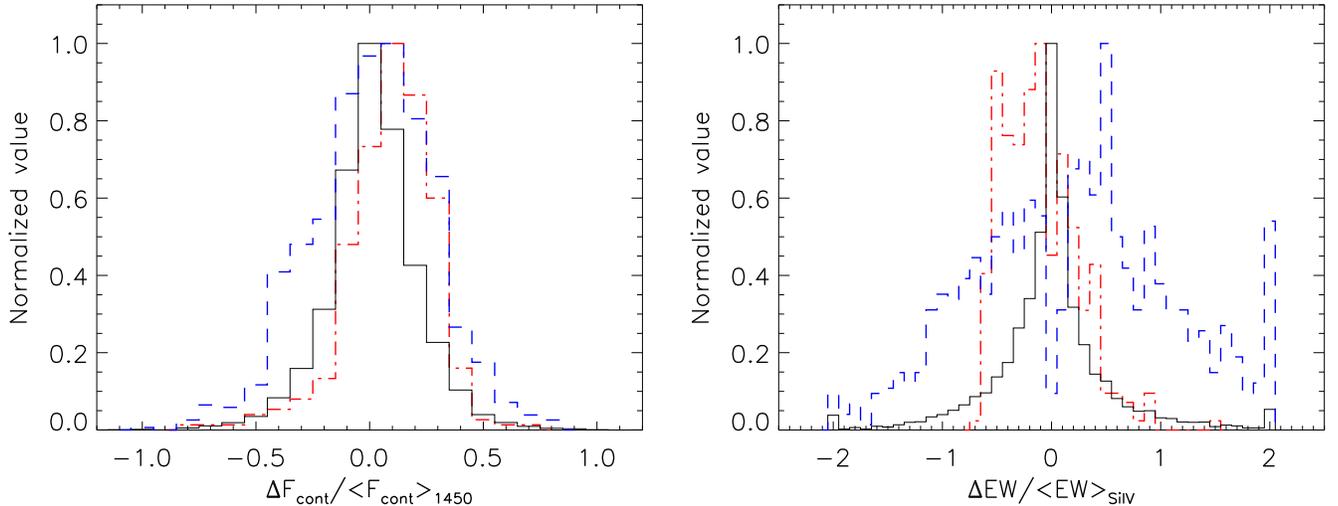}
\caption{{$\Delta \rm F_{\rm cont}/\langle \rm F_{cont} \rangle_{1450}$ (left panel) and $\Delta \rm EW/\langle EW \rangle_{Si\,{\footnotesize IV}}$ (right panel) distributions for the Total sample (solid black), the LLQ2018 sample (dashed blue) and the Phenomenon I sample (dotted-dashed red). Each histogram is normalized to their corresponding maximum value. KS test results indicate that the distributions of $\Delta \rm F_{\rm cont}/\langle \rm F_{cont} \rangle_{1450}$ and $\Delta \rm EW/\langle EW \rangle_{Si\,{\footnotesize IV}}$ show significant ($P\textless1\rm E-8$) differences between the three samples.}
\label{fig.5}}
\end{figure*}

\subsection{{Correlation between fractional variations of Si\,{\footnotesize IV} BALs and the continuum}}\label{subsec:Correlation}

{We plotted the fractional variation of Si\,{\footnotesize IV} BALs ($\Delta \rm EW/\langle EW \rangle_{Si\,{\footnotesize IV}}$) versus the fractional variation of the continuum ($\Delta \rm F_{\rm cont}/\langle \rm F_{cont} \rangle_{1450}$) in Figure \ref{fig.4}. The moderate anticorrelation between them was confirmed by the Spearman rank correlation analysis ($r=-0.51, P=3.20\rm E-22$). This anticorrelation is an important finding to understand the variation mechanism on Phenomenon I.} In fact, Phenomenon I has been mentioned in previous works (e.g., \citealp{Misawa2014,Wang2015,McGraw2018}). \citet{Misawa2014} and \citet{McGraw2018} held the view that the IC scenario was the most plausible explanation for Phenomenon I. 
\citet{Wang2015} pointed out that there {may not be a physical} difference between the variable and nonvariable portions of an absorption trough, {because there may} exist some absorption variations that we cannot detect, even {for} those caused by moderate variability in the ion column density. 
{However, {in previous work, there has been} no convincing evidence to demonstrate the main driver of Phenomenon I, {because no one has studied the behavior of cases where Si\,{\footnotesize IV} absorption varies but C\,{\footnotesize IV} absorption does not.} Importantly, our statistical results provide evidence for the ubiquitous effect of the ionizing continuum variability on the Phenomenon I. }

We attribute the relative lack of variation of the C\,{\footnotesize IV} BALs to the saturation effect. Nonetheless, moderate anticorrelation exits between the fractional variations of the ionizing continuum and Si\,{\footnotesize IV} BALs, indicating that {most of the absorbers are not very optically thick in Si\,{\footnotesize IV} since a highly saturated trough should not show fractional EW variations in response to ionizing continuum variability.} {The range of variation seen for Si\,{\footnotesize IV} in Figure \ref{fig.4} is consistent with that seen by Filiz Ak et al. (\citeyear{Filiz2014}; their Figure 14), when the fractional C\,{\footnotesize IV} EW variation limit is considered.}

{Before this paper, anticorrelation between the fractional flux variations of the continuum and fractional EW variations of both C\,{\footnotesize IV} and Si\,{\footnotesize IV} BALs has been confirmed for a quasar sample (1014 spectrum pairs in 483 quasars, hereafter the LLQ2018 sample) that shows significant variations for both C\,{\footnotesize IV} and Si\,{\footnotesize IV} BALs (\citetalias{Lu2018a}). The phenomenon I sample is the expansion and extension of the LLQ2018 sample, and the anticorrelations presented in these two papers have the consistency in essence. The difference between these two samples is that the BALs in the Phenomenon I sample suffer from more saturations {in C\,{\footnotesize IV}} than that in the \citetalias{Lu2018a} sample.}

{In summary, the correlation between fractional variations of Si\,{\footnotesize IV} BALs and the continuum for the Phenomenon I sample confirms that these BALs are indeed suffering from saturations {in C\,{\footnotesize IV}}, but the IC scenario is still a main driver for their variations {in Si\,{\footnotesize IV}.}

\subsection{Distributions of fractional variations of Si\,{\footnotesize IV} BALs and the continuum}\label{subsec:Distribution}
{It is shown in Figure~\ref{fig.4} that most of the data points have a positive $\Delta \rm F_{\rm cont}/\langle \rm F_{cont} \rangle_{1450}$ value but a negative $\Delta \rm EW/\langle EW \rangle_{Si\,{\footnotesize IV}}$ value. To investigate the cause of this phenomenon, we plotted $\Delta \rm F_{\rm cont}/\langle \rm F_{cont} \rangle_{1450}$ and $\Delta \rm EW/\langle EW \rangle_{Si\,{\footnotesize IV}}$ distributions for the Total sample, the \citetalias{Lu2018a} sample, and the Phenomenon I sample (Figure \ref{fig.5}), respectively. KS tests show that the $\Delta \rm F_{\rm cont}/\langle \rm F_{cont} \rangle_{1450}$ and $\Delta \rm EW/\langle EW \rangle_{Si\,{\footnotesize IV}}$ distributions for the three samples are significantly ($P\textless1\rm E-8$) different from each other.}

{Although the $\Delta \rm F_{\rm cont}/\langle \rm F_{cont} \rangle_{1450}$ distributions show no obvious deviation from symmetry about zero for both the Total sample (with the center $\mu=0.017$ for the best-fitting Gaussian component) and the \citetalias{Lu2018a} sample (with the center $\mu=0.029$ for the best-fitting Gaussian component), the absolute values of $\Delta \rm F_{\rm cont}/\langle \rm F_{cont} \rangle_{1450}$ for the \citetalias{Lu2018a} sample (with the standard deviation $\sigma = 0.265$ for the best-fitting Gaussian component) tend to be larger than that for the Total sample (with the standard deviation $\sigma = 0.156$ for the best-fitting Gaussian component). This indicates that the sample showing BAL variations tends to have larger ionizing continuum variations. Such a phenomenon is reasonable when considering that change of the ionizing continuum as the primary driver of BAL variability (\citealp{Wang2015, He2017}; \citetalias{Lu2018a}). 

{The distributions of $\Delta \rm F_{\rm cont}/\langle \rm F_{cont} \rangle_{1450}$ and $\Delta \rm EW/\langle EW \rangle_{Si\,{\footnotesize IV}}$ for the Phenomenon I sample show deviations from symmetry about zero in the opposite direction  (with the center $\mu=0.116$ and $\mu=-0.185$ of the Gaussian for $\Delta \rm F_{\rm cont}/\langle \rm F_{cont} \rangle_{1450}$ and $\Delta \rm EW/\langle EW \rangle_{Si\,{\footnotesize IV}}$ distributions, respectively). We attribute this trend to the saturation effect on the photoionization-driven BAL variability. An important feature of the Phenomenon I sample is that their C\,{\footnotesize IV} BALs are generally suffering from saturations. This feature is also reflected in the $\Delta \rm EW/\langle EW \rangle_{Si\,{\footnotesize IV}}$ distribution. Because the fractional EW variation measurements can reflect only the lower limits of the optical depth and column density variations for saturated troughs, the $\Delta \rm EW/\langle EW \rangle_{Si\,{\footnotesize IV}}$ distribution of the Phenomenon I sample (with the standard deviation $\sigma = 0.362$ for the best-fitting Gaussian component) tends to have smaller absolute values than the \citetalias{Lu2018a} sample (with the standard deviation $\sigma = 1.042$ for the best-fitting Gaussian component). Since the anticorrelation between the fractional variations of BALs and the continuum has been built (\citetalias{Lu2018a}; Section \ref{subsec:Correlation}), the strengthening of BALs are expected when the ionizing continuum is weakening. However, on the one hand, the EW for a saturated BAL may be unable to be enhanced when its ionization level changes. On the other hand, the strengthening of the ionizing continuum may weaken a saturated BAL if it is not very optically thick. Thus the distributions of the $\Delta \rm F_{\rm cont}/\langle \rm F_{cont} \rangle_{1450}$ and $\Delta \rm EW/\langle EW \rangle_{Si\,{\footnotesize IV}}$ of the Phenomenon I sample show deviations from symmetry of about zero in the opposite direction, which provides statistical evidence for two physical processes {in some saturated BALs:} (1) photoionization-driven transition from saturated to unsaturated and (2) recombination-driven column density strengthening with no significant EW variation.}

\section{Implications of the saturation effect} \label{sec:implication}
\subsection{{Dispersion of the relations between fractional variations of the continuum and BALs}}\label{subsec:dispersion}

The effect of saturation on BALs can well interpret many phenomena of BAL variation that have been discovered previously. For example, \citetalias{Lu2018a} found that although correlations with high significance level are detected, relations between the fractional flux variations of the continuum and fractional EW variations for both Si\,{\footnotesize IV} and C\,{\footnotesize IV} BALs show large dispersion (hereafter Trend I). {Such dispersion is also confirmed in the Phenomenon I sample (see Figure \ref{fig.4}). }We attribute partially this dispersion to the effect of BAL saturation, because the measurements of the fractional EW variation cannot accurately reflect the fractional variations of the optical depth and column density for saturated troughs, they reflect only their lower limits \citep{Arav1999a,Arav2005}.
As in the case of J1537+5829, the EW of the C\,{\footnotesize IV} BAL shows no obvious variation due to the effect of saturation, so the fractional EW variation of the C\,{\footnotesize IV} BAL trough cannot reflect {the true values of the fractional variation of optical depth and column density}.

\subsection{{Larger fractional EW variations for Si\,{\footnotesize IV} than C\,{\footnotesize IV} BALs} }\label{subsec:larger EW}
Previous works have found that Si\,{\footnotesize IV} BALs generally show larger fractional EW variations than C\,{\footnotesize IV} BALs (hereafter Trend II, e.g., \citealp{Filiz2013}, \citetalias{Lu2018a}). Just as \citetalias{Lu2018a} pointed out, Trend II can be attributed to the different fundamental parameters from atomic physics, typically fine structure and oscillator strength, between Si\,{\footnotesize IV} and C\,{\footnotesize IV} (see also section 3.3 of \citetalias{Lu2018b} for a detail description). {In fact, the Phenomenon I is the extreme of Trend II, rather than a qualitatively distinct
phenomenon.} The universality of Trend II also reveals the universality of the saturation in BALs. Based on the above explanations of Trend II, {we find that the EW measurements of Si\,{\footnotesize IV} BAL are statistically more accurate than C\,{\footnotesize IV} BAL when reflecting the true values of optical depth and column density}. So it is not hard to understand why the correlation coefficient between the fractional flux variations of the continuum and fractional EW variations of Si\,{\footnotesize IV} are larger than that of C\,{\footnotesize IV} (table 1 in \citetalias{Lu2018a}). As in J1537+5829, a typical case showing Trend II, the Si\,{\footnotesize IV} BAL shows a fractional EW variation of $-0.35\pm0.010$, while the C\,{\footnotesize IV} BAL shows a fractional EW variation of only $-0.07\pm0.006$ (see Table \ref{tab.1}).

\subsection{{Larger fractional EW variations in weaker BAL troughs}}\label{subsec:weaker BAL}

Previous studies have found that Si\,{\footnotesize IV} and C\,{\footnotesize IV}  BAL variations usually occur in relatively weak troughs, and that weak troughs tend to have larger fractional EW variations than strong troughs (hereafter Trend III; e.g., \citealp{Lundgren2007,Gibson2008,Capellupo2011,Filiz2013,McGraw2018}). 
Trend III is understandable {if weaker absorption troughs are less saturated. If} the weaker absorption troughs tend to suffer less saturation than the stronger ones, {then} the weaker ones would show more variations when responding to the same changes in the ionization condition. Of course, previous studies also found that the absorption troughs in high velocity tend to be weaker and more variable (e.g., \citealp{Gibson2009a,Capellupo2011,Filiz2012,Filiz2013}). Those weaker troughs in higher outflow velocity may have a different origin from the stronger troughs in lower outflow velocity, thus Trend III may be caused by a velocity effect. However, \citet{Filiz2013} demonstrated that the trough weakness rather than a velocity effect is the main driver of Trend III, {by taking the Spearman test on the fractional EW variation versus both average EW over the two relevant epothes and outflow velocity} of C\,{\footnotesize IV} BALs in 291 quasars (see  fig.18 of \citealp{Filiz2013}). Thus, Trend III may attribute mostly to the saturation effect. 

\subsection{{BAL trough often shows only partial variations}}\label{subsec:partial}

Previous studies have showed that variability often occurs only in a portion of a BAL trough (hereafter Trend IV; e.g., \citealp{Gibson2008,Capellupo2011,Capellupo2012,Capellupo2013,Filiz2013}). Some authors tend to believe that Trend IV fits the TM scenario more naturally than the IC scenario (e.g., \citealp{Capellupo2012,Capellupo2013}). We do not rule out the possibility that the TM scenario may account for Trend IV. {However, we point out that the saturation effect on photoionization-driven Type N BAL variability is also easy to generate Trend IV. From the point of view of Type N BAL, a BAL shows saturation feature in large portions does not mean that all NAL systems inside this BAL are saturated. Just like the case of J1537+5829, though, the saturated NAL components (S1$\thicksim$S4) inside C\,{\footnotesize IV} BAL are blended severely, resulting in a typical trough-like profile, the other two weaker and relatively high-speed NAL components (S5 and S6) inside the same trough still show time variations, which can be explained by the IC scenario. Also, the undetectable EW variation of the saturated NAL components (S1$\thicksim$S4) does not mean that there is no change in their ionization level and column density.} Based on the above discussions, we argue that Trend IV does not always mean the TM scenario is the driver of the variability; instead, the IC scenario can also be responsible for Trend IV.

\section{CONCLUSION} \label{sec:conclu}
{We have defined Phenomenon I, which shows significant variation in Si\,{\footnotesize IV} BAL but no, or only small, change in C\,{\footnotesize IV} BAL. The study of this interesting phenomenon helps us understand the role of the saturation feature plays in BAL variation and the physical mechanism behind the variation of saturated BAL. We have shown a typical case of a Type N BAL quasar showing Phenomenon I and have presented a statistical analysis based on multiobserved quasars from SDSS-I/II/III.} Below we summarize our work.

\begin{enumerate}
\item {Based on the two-epoch spectra of J1537+5829, we successfully identify four NAL systems (S1$\thicksim$S4) within its Si\,{\footnotesize IV} BAL and two additional NAL systems (S5 and S6) within its C\,{\footnotesize IV} BAL. The coordinated weakening of all of these NAL systems and the obvious strengthening of the quasar continuum collectively provide evidence supporting the ICs in response to the continuum variations to explain the BAL variability in J1537+5829. We ascribe the none variation of the S1$\thicksim$S4 NAL systems in the C\,{\footnotesize IV} BAL trough to their saturation.} See Section \ref{sec:J1537}.

\item Our statistical analysis based on 74 multiobserved quasars showing Phenomenon I confirms the moderate anticorrelation between the fractional variations of Si\,{\footnotesize IV} BALs and the continuum, revealing the ubiquitous effect of the ionizing continuum variability on Phenomenon I. {Note that in Phenomenon I, the C\,{\footnotesize IV} BAL is strongly saturated but the Si\,{\footnotesize IV} BAL is not.} We attribute the relative lack of variation of C\,{\footnotesize IV} BALs in Phenomenon I to the saturation effects. Besides, the above moderate anticorrelation indicates that these absorbers are not very optically thick {in Si\,{\footnotesize IV}}, and the ICs in response to the continuum variations is still the main driver of their variations. See Section \ref{subsec:Correlation}.

\item {The absolute values of $\rm F_{\rm cont}/\langle \rm F_{cont} \rangle_{1450}$ for the LLQ2018 sample tend to be larger than that for the Total sample. This indicates that the sample showing BAL variations tends to have larger ionizing continuum variations, which serve as a clear result in support of the ionizing continuum variations being responsible for many BAL variations. See Section \ref{subsec:Distribution}.}

\item {The $\rm F_{\rm cont}/\langle \rm F_{cont} \rangle_{1450}$ and $\Delta \rm EW/\langle EW \rangle_{Si\,{\footnotesize IV}}$ distributions for the Phenomenon I sample show deviations from symmetry of about zero in the opposite direction. This can be explained by the saturation effect on the photoionization-driven BAL variability.} See Section \ref{subsec:Distribution}.

\item {Finally, we demonstrate that the saturation effect on BAL variability can well explain many phenomena of BAL variations that have been reported before.} See Section \ref{sec:implication}.

\end{enumerate}

\acknowledgments
We are very grateful to the anonymous referee for many comments that greatly improved the quality of this article. In addition, we thank He et al. (\citeyear{He2017}) for making the multiobserved BAL catalog publicly available.

Funding for SDSS-III was provided by the Alfred P. Sloan Foundation, the Participating Institutions, the National Science Foundation, and the US Department of Energy Office of Science. The SDSS-III website is \url{http://www.sdss3.org/}.

SDSS-III is managed by the Astrophysical Research Consortium for the Participating Institutions of the SDSSIII Collaboration, including the University of Arizona, the Brazilian Participation Group, Brook haven National Laboratory, Carnegie Mellon University, University of Florida,the French Participation Group, the German Participation Group, Harvard University, the Instituto de Astrofisica deCanarias, the Michigan State/Notre Dame/JINA Participation Group, Johns Hopkins University, Lawrence Berkeley National Laboratory, Max Planck Institute for Astrophysics, Max Planck Institute for Extraterrestrial Physics, New Mexico State University, New York University, Ohio State University, Pennsylvania State University, University of Portsmouth, Princeton University, the Spanish Participation Group, University of Tokyo, University of Utah,Vanderbilt University, University of Virginia, University of Washington, and Yale University.

\bibliographystyle{aasjournal}
\bibliography{NALvsBAL} 

\end{document}